\documentclass[aps,pre,twocolumn,superscriptaddress]{revtex4}

\usepackage{color}
\usepackage{graphicx}
\usepackage{ulem}
\usepackage{bm}

\begin{document}



\title{Generalized Continuous Maxwell Demons}


\author{Juan P. Garrahan}
\affiliation{School of Physics and Astronomy, University of Nottingham, Nottingham, NG7 2RD, UK}
\affiliation{Centre for the Mathematics and Theoretical Physics of Quantum Non-Equilibrium Systems,
University of Nottingham, Nottingham, NG7 2RD, UK}

\author{Felix Ritort}
\affiliation{Small Biosystems Lab, Condensed Matter Physics Department, Universitat de Barcelona, C/ Mart\'{\i} i Franqu\`es 1, E-08028, Barcelona (Spain)}


\begin{abstract}
\noindent 
We introduce a family of Generalized Continuous Maxwell Demons (GCMDs) operating on idealized single-bit equilibrium devices that combine the single-measurement Szilard and the repeated measurements of the Continuous Maxwell Demon protocols. We derive the cycle distributions for extracted work, information-content, and time and compute the power and information-to-work efficiency fluctuations for the different models. We show that the efficiency at maximum power is maximal for an opportunistic protocol of continuous type in the dynamical regime dominated by rare events. 
We also extend the analysis to finite-time work extracting protocols by mapping them to a three-state GCMD. We show that dynamical finite-time correlations in this model increase the information-to-work conversion efficiency, underlining the role of temporal correlations in optimizing information-to-energy conversion. The effect of finite-time work extraction and demon memory resetting is also analyzed. We conclude that GCMD models are thermodynamically more efficient than the single-measurement Szilard and preferred for describing biological processes in an information-redundant world.
\end{abstract}

\maketitle

\section{Introduction}
\label{sec:intro}
Information-to-work conversion is a fundamental process in physics and biology. Paradigmatic examples are the Maxwell demon, and the Szilard engine, small operating devices that fully convert heat into work using measurement information \cite{Leff1990,Bennett1982,Lutz2015}. According to Landauer and Bennett such devices do not violate the second law as the erasure procedure required to restore the system's initial state increases the overall entropy offsetting Demon's gain \cite{maruyama2009}. The field of thermodynamics of information is witnessing major progress \cite{Sagawa2010,Sagawa2014,Parrondo2015} as these ideas are expanding into new directions \cite{strasberg2013,roldan2014,koski2015,martinez2016,pekola2019}, being also experimentally tested \cite{Toyabe2010,berut2012,Koski2014,Vidrighin2016,Gavrilov2016,Chida2017,Admon2018,Kumar2018,Paneru2018,Manzano2021}. 

Recently, a continuous version of the Maxwell demon (CMD) has been introduced and implemented in a Szilard information-to-work engine operating on single DNA hairpins \cite{Ribezzi2019, Ribezzi2019b}. The CMD offers a view complementary to the standard Szilard engine that is analytically solvable and can be implemented experimentally. In the Szilard engine (hereafter referred to as SZ), a gas particle occupies a vessel with two compartments in contact with a thermal bath at temperature $T$. At a given time, an observation is made of the compartment occupied by the particle. A work extraction process follows by inserting a pulley mechanism with a movable wall such that the compartment reversibly expands under the particle collisions. The combination of measurement, work extraction, and demon resetting defines a cycle in the SZ engine. For a given compartment, the average extracted work per cycle equals $-k_BT\log p$ where $p$ is the probability of observing the particle in that compartment. For two compartments of probability $p_0,p_1$ the average extracted work per cycle in the SZ engine equals $W_{\rm SZ}=-k_BT(p_0\log p_0+p_1\log p_1)$ which is bounded from above by $W_{\rm SZ}\le k_BT\log 2$ for the case $p_0=p_1=1/2$. For an irreversible work extraction protocol, the average work extracted is further bounded by the information-content of the single-bit measurement expressed in nats, $I_{\rm SZ}=-(p_0\log p_0+p_1\log p_1)$ also called the Landauer limit, $W\le k_BT I$. The CMD setting is the same as for SZ; the only difference is the work extraction condition and the cycle length. In the CMD, an observation of the compartment occupied by the particle is made, but work is not extracted right away. Instead, measurements are repeatedly made every time $\tau$ from the initial observation until a change in the compartment occupied by the particle is observed. Note that in the CMD a decision to extract work is taken based on a series of observations made at every $\tau$-consecutive measurement, irrespective of any unobserved transition in between. Therefore, any cycle in the CMD contains at least two bits (for the first compartment observation and the next observed compartment change) rather than the single-bit cycle of SZ. The information-content of the multiple-bit stored sequences of the CMD cycle is always larger than for SZ, permitting the extraction of more work per cycle in the former. The average work per cycle extracted in the CMD is $\tau$-independent and given by, $W_{\rm CMD}=-k_BT(p_0\log p_1+p_1\log p_0)$. In contrast to SZ, $W_{\rm CMD}$ is now bounded from below by $k_BT\log 2$, $W_{\rm CMD}\ge k_BT\log 2$, with $W_{\rm CMD}=k_BT\log 2$ for $p_0=p_1=1/2$. The average work per cycle is bounded from above by the information-content of the multiple-bit sequences generated by the CMD, $W_{\rm CMD}\le k_BT I_{\rm CMD}(\tau)$ where $\tau$ is the time between consecutive measurements. For uncorrelated  observations ($\tau\gg 1/R$ with $R$ systems' relaxation rate), the information-content of the stored sequences is minimal and equals $I_{\rm CMD}(\tau\to\infty)=-(p_0\log p_0)/p_1-(p_1\log p_1)/p_0+W_{\rm CMD}/k_BT$ defining the Landauer limit in this case, $W_{\rm CMD}< k_BT I_{\rm CMD}$. Given the simplicity of the CMD protocol and its complementarity to SZ, it is surprising that the CMD was not conceived before. The motivation in Ref. \cite{Ribezzi2019} was to devise a model for which large amounts of work could be extracted. The average work extracted $-k_BT\log p$ is large for compartments of low $p$, so a measurement protocol designed to seek for low $p$ events maximizes work extraction. Indeed, in the limits $p_0\to 0$ and $p_0\to 1$, $W_{\rm CMD}$ diverges while $W_{\rm SZ}$ vanishes.

In the single-molecule experimental realization of the SZ and CMD, a DNA hairpin hops between two states, folded and unfolded, of probabilities $p_0$ and $p_1=1-p_0$, and a discrete or continuous feedback protocol was implemented to extract work depending on the measurement outcome \cite{Ribezzi2019,Ribezzi2019b}. Work extraction is operated at the first measurement time in the SZ engine. In contrast, in the CMD, measurements are repeatedly made at intervals $\tau\sim 10^{-3}s$, and work is extracted only when the system is observed to switch state for the first time (folded to unfolded or unfolded to folded). As previously said,the CMD exhibits novel features as compared to the SZ, such as a larger average work per cycle (with the Landauer limit being a lower bound rather than an  upper bound) and maximum efficiency $\eta<1$ in the regime dominated by rare transition events ($p_0\to 0,1$). This paper investigates power fluctuations in the CMD, establishing the fundamental differences between these two classes (discrete versus continuous) of information machines. The paper has three main parts. In Section \ref{sec:GCMD} we introduce the generalized continuous Maxwell demon (GCMD) as a two-pathway model for information-to-energy conversion that combines features of SZ and CMD. Section \ref{sec:fluctuations} analyzes cycle-power and cycle-efficiency fluctuations in the GCMD, the main distinctive feature of continuous models compared to the SZ model. Section \ref{sec:finitetime} addresses the temporal correlations introduced in Ref.\cite{Admon2018} on the power and efficiency at finite times. Finally, Sections \ref{sec:discussion} and \ref{sec:conclusions} are devoted to discussion and conclusions.
 \begin{figure}[t!] 
   \centering
   \includegraphics[width=\columnwidth]{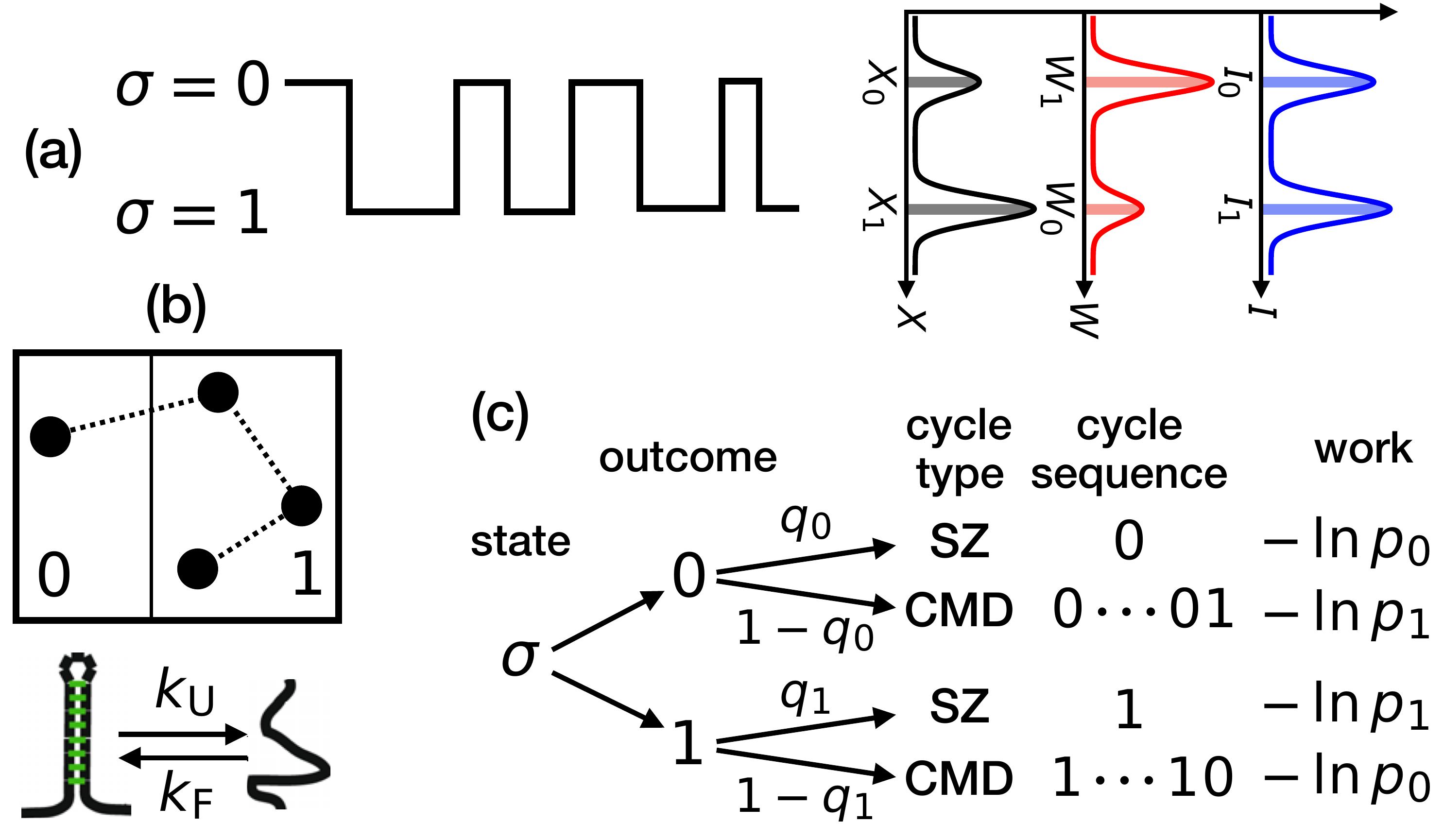} 
   \caption{{\bf Illustration of the GCMD.} (a) Single-bit dichotomous signal (left) and position, work and information distributions (right). In a noisy environment fluctuations widen the single-peaked distributions into Gaussians. (b) Two examples of the model: a single particle moving in a compartmentalized vessel and a molecule hopping between two states. (c) Schematics of the work-extracting protocol in the GCMD. (SZ: Szilard-type protocol; CMD: Continuous-type protocol).}
   \label{fig:FIG1}
\end{figure}
\section{The generalized continuous Maxwell demon (GCMD) model}
\label{sec:GCMD}
We introduce a new family of generalized continuous Maxwell demons (GCMD) that expands the previously studied work extraction protocols to multiple repeated measurements and has the SZ and the CMD as particular cases. Figure~1 illustrates the GCMD in the case of a single-bit measurement outcome (0,1). 

A GCMD operates as follows. A system hops between two states $\sigma=0,1$ generating a two-level dichotomous signal (Fig.~\ref{fig:FIG1}a) of probability $p_\sigma$ ($p_0+p_1=1$). Examples are: (i) a free molecule in a volume $V$ that is subdivided into two compartments ($V_0,V_1$) such that $V=V_0+V_1$; (ii) a biomolecule in solution with two conformations, folded ($\sigma=0$) and unfolded ($\sigma=1$). For a system in equilibrium, we have $p_0=1/(1+\exp(\Delta G/k_BT))$ with $\Delta G$ the free energy difference between states 0 and 1 (hereafter, we set $k_BT=1$). In example (a), $\Delta G= \ln(V_1/V_0)$ and $p_\sigma=V_\sigma/V$ (Fig.~\ref{fig:FIG1}b,top). In example (b),  $\Delta G$ is the folding free energy and $p_0/p_1=\exp(-\Delta G)$ (Fig.~\ref{fig:FIG1}b,bottom). A cycle in the GCMD starts with an observation of the system's state $\sigma$ and two possible actions: with probability $q_\sigma$ the SZ protocol is operated, and the amount of work per cycle ($W^{\rm SZ}_{\sigma}=-\log p_\sigma$) readily extracted; with probability $1-q_\sigma$ the CMD protocol is operated, meaning that new observations are made every $\tau$ until the system switches state $\sigma\to 1-\sigma$ and the work per cycle ($W^{\rm CMD}_{\sigma}=-\log p_{1-\sigma}$) extracted. The $q_0,q_1$ define two independent processes with $0\le q_0,q_1\le 1$ (i.e.
$q_0+q_1$ must not be equal to 1). The model interpolates between SZ ($q_0=q_1=1$) and CMD ($q_0=q_1=0$). The GCMD model is symmetric with respect to the transformation $p_0\leftrightarrow 1-p_0$ and $0\leftrightarrow 1$, so we can restrict the analysis to $0<p_0\le 1/2$.   

\subsection{Average work, information-content, and cycle-time}
\label{subsec:WIT}
The average work per cycle in the GCMD (in $k_BT$ units) is given by,
\begin{eqnarray}
    \overline{W}&=&\sum_{\sigma=0,1}p_\sigma\bigl(q_\sigma W^{\rm SZ}_{\sigma}+ (1-q_\sigma)W^{\rm CMD}_{\sigma}\bigr)  \label{work1} 
    \\
    & =& -\sum_{\sigma=0,1}p_\sigma\bigl(q_\sigma  \ln p_\sigma+ (1-q_\sigma) \ln (1-p_\sigma)\bigr)\,\,.
    \nonumber
\end{eqnarray}

Similarly, we analyze the information-content of the stored sequences of cycles ${\cal C}$, defined as $I({\cal C})=- \ln P({\cal C})$ (nat units). SZ cycles are one-bit sequences, ${\cal C}=\lbrace \sigma \rbrace$, with $P({\cal C})=p_\sigma q_\sigma$. 
In contrast CMD cycles consist of $n+1$ ($n\ge 1$) bit-sequences that start at the first bit ($\sigma$) which is repeated $n$ times, until the bit outcome switches ($\sigma\to 1-\sigma$) at the $(n+1)^{\rm th}$ time . Therefore, a CMD cycle contains at least two-bits, ${\cal C}=\lbrace \overbrace{\sigma,\sigma,...,\sigma}^{n},1-\sigma
\rbrace$, where $1-\sigma$ indicates state switching.
Note that the stopping time $n$ is stochastic, and varies from cycle to cycle. The probability of the CMD cycle is given by $P({\cal C})=p_\sigma(1-q_\sigma)T_{\sigma \sigma}^{n-1}(1-T_{\sigma\sigma})$, where $T_{\sigma\sigma}$ is the conditional probability of a repeated measurement outcome $\sigma$ after time $\tau$. It has been shown \cite{Ribezzi2019,Ribezzi2019b} that the lowest sequence information-content (i.e.\ minimum redundancy) is obtained for fully uncorrelated bit sequences, that is, when the relaxation kinetic rate of the system $R$ is such that $R\tau\gg 1$, in which case $T_{\sigma\sigma}=p_\sigma$. In this limit, the probability of a GCMD cycle is given by,
\begin{eqnarray}
P_{\rm GCMD}({\cal C}) = \left\{
\begin{array}{cl}
 q_\sigma p_\sigma    & (n=1, \text{SZ}) \\
 (1-q_\sigma) p_{1-\sigma} p_{\sigma}^{n}  & (n\geq1, \text{CMD}) \\
\end{array}
\right. ,
\label{PC}
\end{eqnarray}
and the corresponding average information-content
\begin{eqnarray}
    \overline{I}&=&\sum_{{\cal C}}P({\cal C})I({\cal C})=-\sum_{\sigma=0,1} p_\sigma q_\sigma  \ln(p_\sigma q_\sigma)
    \label{ic1}- \\ 
   &&  \sum_{\sigma=0,1} (1-q_\sigma)p_{1-\sigma}\sum_{n=1}^{\infty}p_\sigma^n  \ln((1-q_\sigma)p_{1-\sigma}p_\sigma^n)
    \nonumber
\end{eqnarray}
where we used $p_{1-\sigma}=1-p_{\sigma}$. 

In the context of the thermodynamics of feedback, the thermodynamic efficiency of information-to-energy conversion is defined as the ratio between the average work extracted ($\overline{W}$) relative to the average energy converted into heat ($\overline{Q}$) that is necessary to erase the stored sequences: $\eta_{\rm th}=\overline{W}/\overline{Q}$. According to the second law, $\overline{W}\le\overline{Q}$ and $\eta_{\rm th}\le 1$.  Following Landauer, the minimum energy $\overline{Q}$ for erasure (assuming a zero-work measurement process) is given by the Shannon information-content $\overline{I}$ of the stored sequences \cite{schmitt2015molecular, Admon2018} as given in Eq.~(\ref{ic1}) in the limit case of decorrelated measurements, $R\tau\gg 1$. Therefore, from Eqs.(\ref{work1},\ref{ic1}) we define the {\it thermodynamic efficiency}  as $\eta_{\rm th}=\overline{W}/\overline{I}$. The second law ensures that $\eta_{\rm th}\leq 1$ which for SZ is saturated ($\eta_{\rm th}=1$) while for CMD, $\eta_{\rm th}>1/3$ approaching 1 in the limit $p_0\to 0$ where rare events dominate dynamics \cite{Ribezzi2019}.

We also consider the thermodynamic power $P_{\rm th}=\overline{W}/\overline{t_{\cal C}}$,
as the ratio of $\overline{W}$, Eq.~(\ref{work1}), and the average cycle time, $\overline{t_{\cal C}}$. Combining SZ cycles (duration $n=1$) and CMD cycles ($n+1$ duration, with $n\geq 1$), we get for $\overline{t_{\cal C}}$ (in $\tau$ units),
\begin{eqnarray}
\overline{t_{\cal C}} &=& \sum_{\sigma=0,1}
\left[
1\cdot p_\sigma q_\sigma  +  (1-q_\sigma)\sum_{n\ge 1}(n+1) p_\sigma^n (1-p_{\sigma})
\right]
\nonumber\\
&=&
1+(1-q_0)\frac{p_0}{p_1}+(1-q_1)\frac{p_1}{p_0}
\label{cycletime}
\end{eqnarray}
which reduces to $\overline{t_{\cal C}}=1$ for SZ ($q_\sigma=1$) and to $\overline{t_{\cal C}}=(p_0(1-p_0))^{-1}-1\ge 3$  \cite{Ribezzi2019} for CMD ($q_\sigma=0$). While the CMD does extract more work than the SZ \cite{Ribezzi2019}, this is at the price of a larger $\overline{t_{\cal C}}$: in both cases, $P_{\rm th}$ vanishes asymptotically like $-p_0 \ln p_0$ when $p_0\to 0$. 
 \begin{figure}[t!] 
   \centering
   \includegraphics[width=\columnwidth]{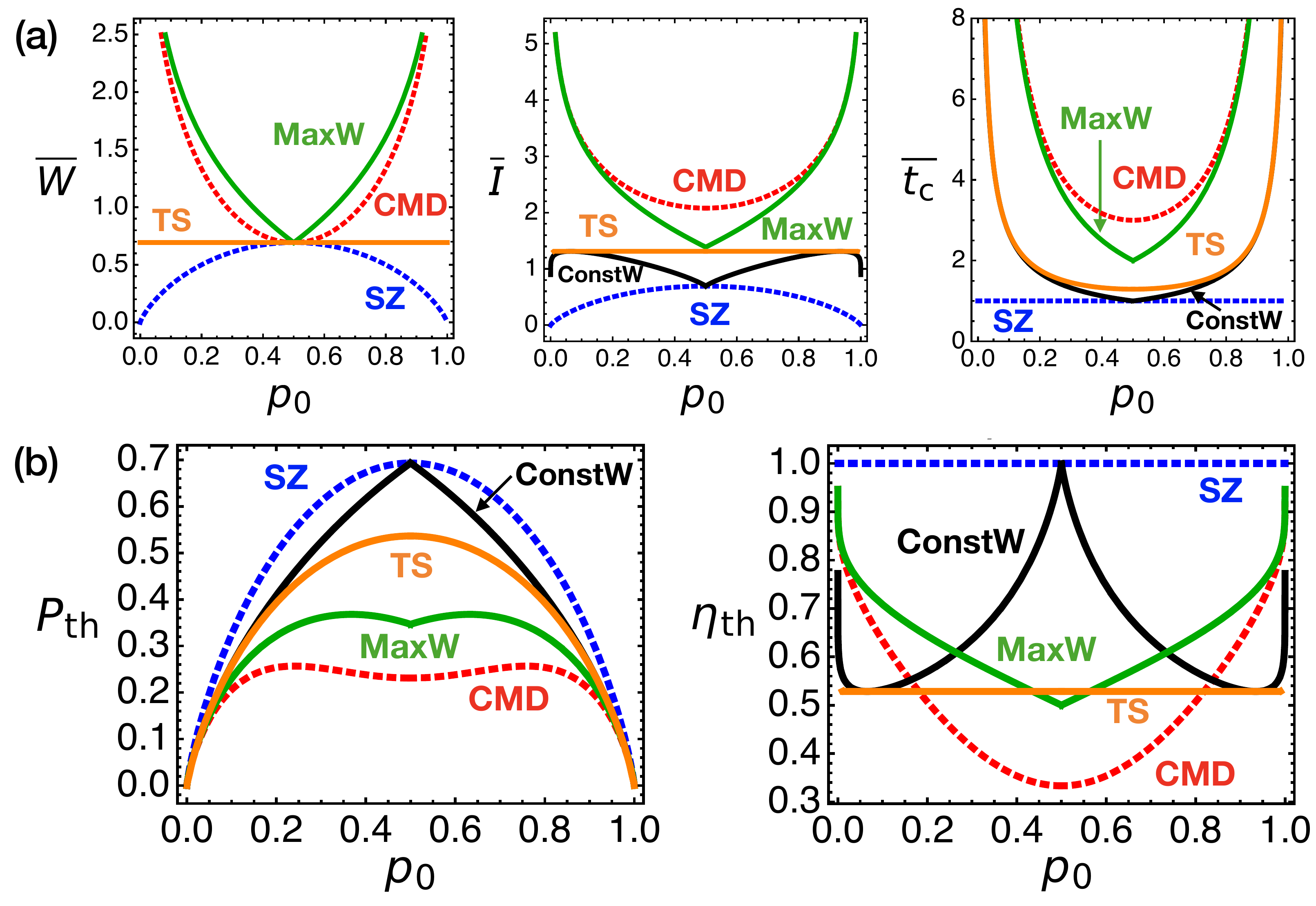} 
   \caption{{\bf Thermodynamic parameters for the different GCMD models.} (a) Average work ($\overline{W}$), information-content ($\overline{I}$) and cycle-time ($\overline{t_{\cal C}}$) versus $p_0$. (b) Average power ($P_{\rm th}$) and efficiency ($\eta_{\rm th}$) versus $p_0$. In the limit $p_0\to 0,1$ all models reach maximum efficiency, $\eta_{\rm th}\to 1$. Note the symmetry $p_0\to 1-p_0$.  (SZ: Szilard model, blue; CMD: Continuous Maxwell Demon, red; MaxW: Maximum work, green; ConstW: Constant work, black; TS: Thermostable, orange).}
   \label{fig:FIG2}
\end{figure}
\subsection{Family of GMCD models}
\label{subsec:specific}
Besides the SZ and CMD particular cases, we now consider other physically relevant choices defined by the $q_{\sigma}$ values. These are: \\
(i) The {\it maximum work model} (MaxW) where $q_0=1,q_1=0$ if $p_0\le p_1$ ($q_0=0,q_1=1$ if $p_0\ge p_1$) which ensures maximum extracted work $\overline{W}=- \ln(\min_{\sigma}(p_{\sigma}))$. This model may be called {\it opportunistic} because it extracts the maximum work for all measurement outcomes.\\
(ii) The {\it constant work model} (ConstW) where $\overline{W}= \ln(2)$, $\forall p_0$. From Eq.~(\ref{work1}) we see that for this to occur $q_0$ and $q_1$ have to obey, 
\begin{equation}
\text{ConstW:} \;\;  p_0q_0+p_1(1-q_1)= \ln(2p_1)/ \ln(p_1/p_0) . \label{CW}
\end{equation}
(iii) The {\it thermostable model} (TS) where both $\overline{W}$ and $\overline{I}$ are independent of $p_0$ in which case $\eta_{\rm th}={\rm const}$. This is obtained by applying Eq.~(\ref{CW}) to solve for $q_1$ in terms of $q_0$ and $p_0$, replacing in $\overline{I}$, and then finding a dependence of $q_0$ on $p_0$ which makes $\overline{I}$ independent of $p_0$ \footnote{There are multiple solutions of $q_{0,1}$ as functions of $p_0$ that give $\overline{W}$, $\overline{I}$ constant. We choose the one that simultaneously spans the largest range of $p_0$ and minimizes $\overline{I}$. This solution exists in the range $0.01 \lesssim p_0 \lesssim 0.99$.}.

Figure \ref{fig:FIG2}  shows $\overline{W},\overline{I},\overline{t_{\cal C}},\eta_{\rm th},P_{\rm th}$ for the different models. We observe that SZ beats all models in terms of $P_{\rm th},\eta_{\rm th}$; however, SZ also gives the lowest average work per cycle (Fig. \ref{fig:FIG2}a). While the models exhibit different features (CMD: lowest $P_{\rm th}$ and largest $\overline{I}$; MaxW: largest $\overline{W}$; ConstW and TS: large $P_{\rm th}$) all GCMD variants show the same trend: $\eta_{\rm th}\to 1$ and $P_{\rm th}\to 0$ when $p_0\to 0,1$.

\subsection{Physical interpretation of the GMCD model}
\label{subsec:physical}
The SZ protocol describes what might be called a {\it store-take strategy} where the most probable state is the one preferentially chosen for work extraction. Instead, in the CMD the first measurement of the system’s state is followed by repeated measurements (store) until a specific condition is met (check) and the work extraction is operated (take). The {\it store-check-take strategy} of the CMD extracts work from the less probable state, maximizing average work extraction compared to SZ. One might call the two strategies low risk (SZ) and high risk (CMD). Here risk stands for repetitive measurement operations that increase the information content of the stored sequences and extract work only at the end. High-risk strategies rely on rare events of low probability $p$ and large information content that deliver a large amount of work, $-k_B T \log p$. High-risk strategies rely on rare events of low probability p and large information content that deliver a large amount of work, $-k_B T \log p$. Instead, low-risk strategies rely on high probability p and low information-content events that deliver a small amount of work, $-k_B T \log p\simeq k_B T(1-p)$ for $p$ close to 1. High-risk strategies (CMD) trade large amounts of work and information, whereas low-risk strategies (SZ) trade small amounts of work and information.

\section{Cycle-power and cycle-efficiency fluctuations}
\label{sec:fluctuations}
A main feature of the GCMD is the stochastic nature of $W, I$ and $t_{\cal C}$, which leads to large fluctuations in the power and efficiency when measured over individual cycles \cite{verley2014},\cite{paneru2020efficiency}. From Eq.~(\ref{PC}) we can readily derive the corresponding $W,I,t_{\cal C}$ distributions from which the distributions for cycle-power $P=W/t_{\cal C}$ (in $k_BT/\tau$ units) and cycle-efficiency $\eta=W/I$ (denoted as cycle-$P$ and cycle-$\eta$) follow. For cycle-$P$ we get a discrete distribution
\begin{eqnarray}
    {\cal P}(P) &=& \sum_{\sigma=0,1}
    \left[ 
    p_\sigma q_\sigma \delta(P+ \ln p_\sigma)        \phantom{\frac{(1-q_\sigma)}{p_\sigma^{1/P}}}
    \right.
    \nonumber\\
    && 
    \left.
    +\frac{p_{1-\sigma}(1-q_\sigma)}{p_\sigma^{1+ (\ln p_{1-\sigma})/P}}  \theta(P_{1-\sigma}^*-P)
    \right] \, ,
    \label{cycle-power}
\end{eqnarray}
where $\theta$ is the Heaviside function, $P^*_\sigma=- (\ln p_\sigma)/2$ are thresholds, and the values that $P$ can take in the argument of the Heaviside functions for each $\sigma$ are discrete:
$P=P^*_{1-\sigma} / (n+1)$ ($n\geq1$). Similarly, for cycle-$\eta$ we get,
\begin{equation}
    {\cal P}(\eta) = \sum_{\sigma=0,1}
    \left[ 
    p_\sigma q_\sigma \delta\left(\eta-
    \frac{\ln p_\sigma}{ \ln(p_\sigma q_\sigma)}\right) + 
    p_{1-\sigma}^{\frac{1}{\eta}}  \theta(\eta_\sigma^*-\eta)
    \right] 
    \label{cycle-efficiency}
    \, ,
\end{equation}
where $\eta_\sigma^*=(1+ \ln(p_{\sigma}(1-q_{\sigma}))/ \ln p_{1-\sigma})^{-1}$ are thresholds, and in the Heaviside functions $\eta$ takes discrete values $\eta=(1+ \ln(p_{\sigma}^n(1-q_{\sigma}))/ \ln p_{1-\sigma})^{-1}$ $(n\geq 1)$.
As expected, SZ ($q_0=q_1=1$) does not fluctuate: ${\cal P}(P)=p_0\delta(P+ \ln p_0)+p_1\delta(P+ \ln p_1)$ and ${\cal P}(\eta)=\delta(\eta-1)$. 

In the large fluctuation regime of continuous-type models ($p_0\to 0$), distributions can be approximated by:
\begin{eqnarray}
  {\cal P}_{p_0\to 0}(P)&\approx&q_1\delta(P)+\frac{p_0(1-q_1)}{(1-p_0)^{1+ (\ln p_0)/P}} , \label{cycle-powerlimit}\\
  {\cal P}_{p_0\to 0}(\eta)&\approx&q_1\delta(\eta)+p_0^{1/\eta} , \label{cycle-etalimit}
\end{eqnarray}
(where the values of $P$ and $\eta$ in the second terms of  the rhs, are discretized as above). 

\begin{figure}[t!] 
   \centering
   \includegraphics[width=\columnwidth]{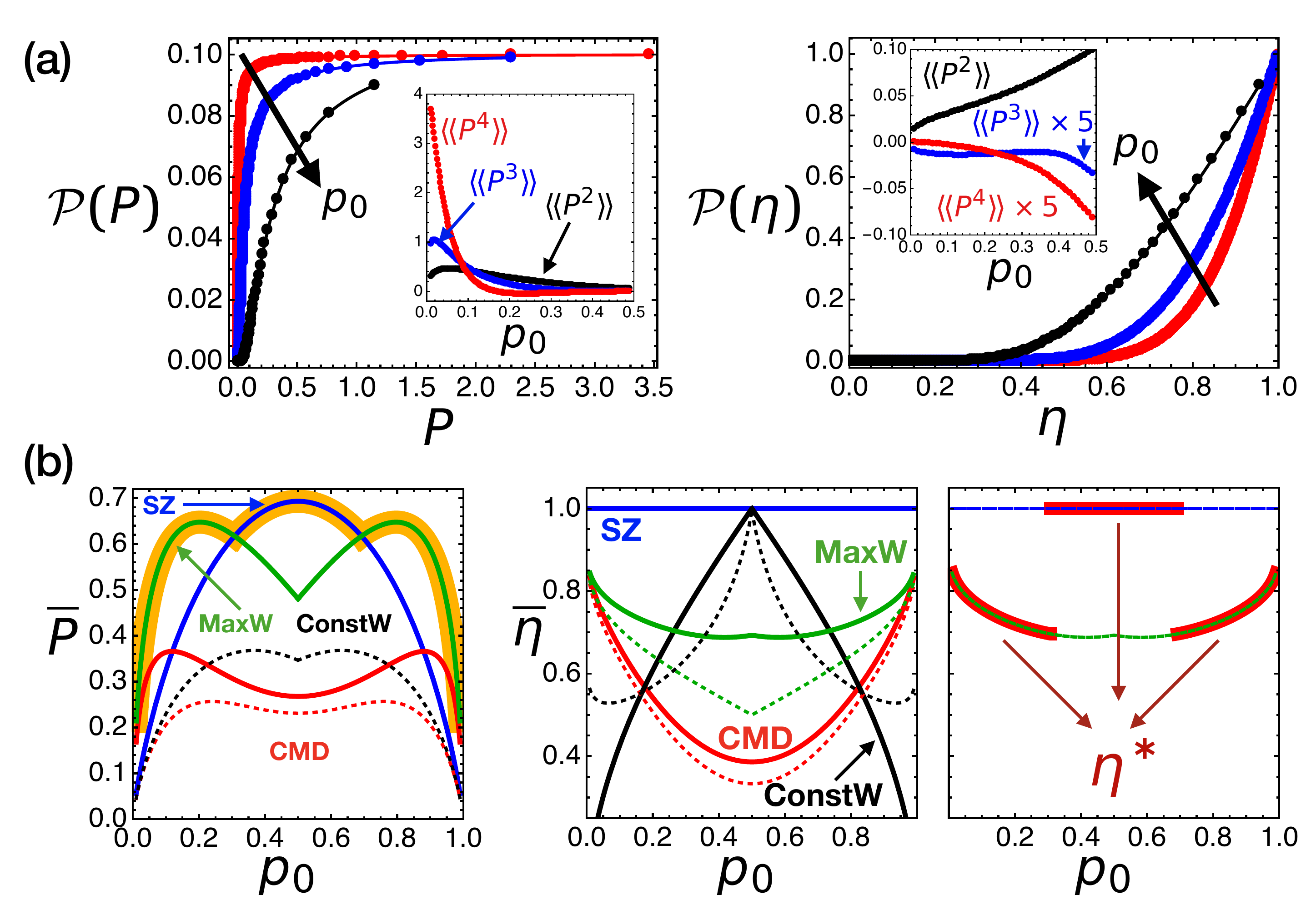} 
   \caption{{\bf Statistics of cycle-power and cycle-efficiency for the different GCMD models.} (a) Cycle-power (left) and cycle-efficiency (right) distributions for decreasing values of $p_0=$0.1 (black), 0.01 (blue), 0.001 (red)  for the MaxW model. Insets show the second, third, and fourth cumulants of the distributions versus $p_0$. (b) Average cycle-power ($\overline{P}$, left) and cycle-efficiency ($\overline{\eta}$, middle) versus $p_0$ (continuous lines). Dashed lines are the thermodynamic efficiencies of Fig.\ref{fig:FIG2}. Orange thick line in the left panel is the maximum power as given by MaxW ($p_0\lesssim 0.3$) and SZ ($0.3\lesssim p_0\le 1/2$). Information-to-work efficiency at maximum power (right panel, red). Models abbreviations and colors as in Fig.\ref{fig:FIG2}. }
   \label{fig:FIG3}
\end{figure}

We find an interesting asymptotic behavior for power and efficiency fluctuations in the small $p_0$ limit. In Fig.~\ref{fig:FIG3}a we show  ${\cal P}(P)$ and  ${\cal P}(\eta)$ for various values of $p_0 \ll 1$ for the MaxW model ($q_0=1,q_1=0$): the probability of the power becomes increasingly uniform with decreasing $p_0$, while that of the efficiency peaks at $\eta=1$. The insets in Fig.~\ref{fig:FIG3}a show the second, third, and fourth cumulants of the distributions as a function of $p_0$ (all cumulants vanishing for SZ). 
In Fig.\ref{fig:FIG3}b we show the first moments of (\ref{cycle-power}) and (\ref{cycle-efficiency}), the average cycle-power ($\overline{P}$) and cycle-efficiency ($\overline{\eta}$) for several models. These are larger than the corresponding thermodynamic values, $\overline{P}\ge P_{\rm th}$ and $\overline{\eta}\ge \eta_{\rm th}$, for a wide range of $p_0$.  

Notice that while $P_{\rm th}$ is always maximum for SZ, $\overline{P}$ is maximum for SZ in the intermediate regime $0.3 \lesssim p_0 \lesssim 0.7$ \footnote{These limits are $x$ and $1-x$, where $x$ is the the solution of the equation 
$(1 - p_0) \ln{(1 - p_0)} + p_0 \ln{p_0} (1 +  \ln{p_0}/(1 - p_0))=0$.}, while in the strongly fluctuating regime $p_0\to 0$ (or $p_1 \to 0$),  $\overline{P}$ is maximum for the MaxW model, see left panel of Fig.~\ref{fig:FIG3}b (orange envelope). Although $\eta_{\rm th}=\overline{\eta}=1$ for SZ, this is at the cost of the lowest $\overline{P}$ in the limit $p_0\to 0$ where GCMDs are most relevant. In the right panel of Fig.~\ref{fig:FIG3}b we show the efficiency at maximum power, $\eta^*$, among all possible models ($q_{\sigma}$). In the intermediate regime $0.3 \lesssim p_0 \lesssim 0.7$ $\eta^*=1$ is maximal for SZ, whereas if $p_0 \lesssim 0.3$ (or $p_1 \lesssim 0.3$) 
we get $\eta^*=\overline{\eta}_{\rm MaxW}$.
These results demonstrate: 1) the efficiency at maximum power is strongly sensitive to fluctuations and; 2) in the rare events regime, $p_0 \lesssim 0.3$, an opportunistic (MaxW) model maximizes the efficiency.

\subsection{{\it Store-take} versus {\it store-check-take} strategies}
\label{subsec:strategies}
One may ask which average (thermodynamical or cycle-average) is physically more relevant. In small biological systems (such as molecular machines operating in a cellular environment), we envision dynamics as a sequence of repeated regulatory-feedback cycles of different time duration. During each regulatory cycle, a physical variable is continuously monitored until a specific condition is met, at which point information-to-energy conversion occurs. For example, neuronal transmission consists of three phases \cite{bialek2012}: first, a stimulus drives cell depolarization and a rise in membrane potential ({\it storage} phase); the membrane potential reaches a threshold value ({\it check} phase); the action potential (spike) is triggered followed by cell repolarization ({\it take} phase). Upon cycle termination, the system is reset, and the stored information is erased. The duration of these elementary cycles cannot be too long as the robustness of the stored information is at stake in the noisy cellular environment. In other words, the transduction and consumption of the energy accumulated in regulatory-feedback cycles inside the cell must occur over time intervals sufficiently short for the stored information to persist before the thermal forces erase it. In a strategy of {\it store-check-take}, it seems more appropriate to consider cycle quantities rather than thermodynamic quantities averaged over long times. The efficiency at maximum power $\eta^*$ in the region $0<p_0\lesssim 0.3$ where continuous GCMD models show a higher power (Fig.~\ref{fig:FIG3}b, right panel) suggests that the {\it store-check-take} strategy of repeated measurement protocols is better than the single-measurement action {\it store-take} of the SZ model.
 \begin{figure}[t!] 
   \centering
   \includegraphics[width=\columnwidth]{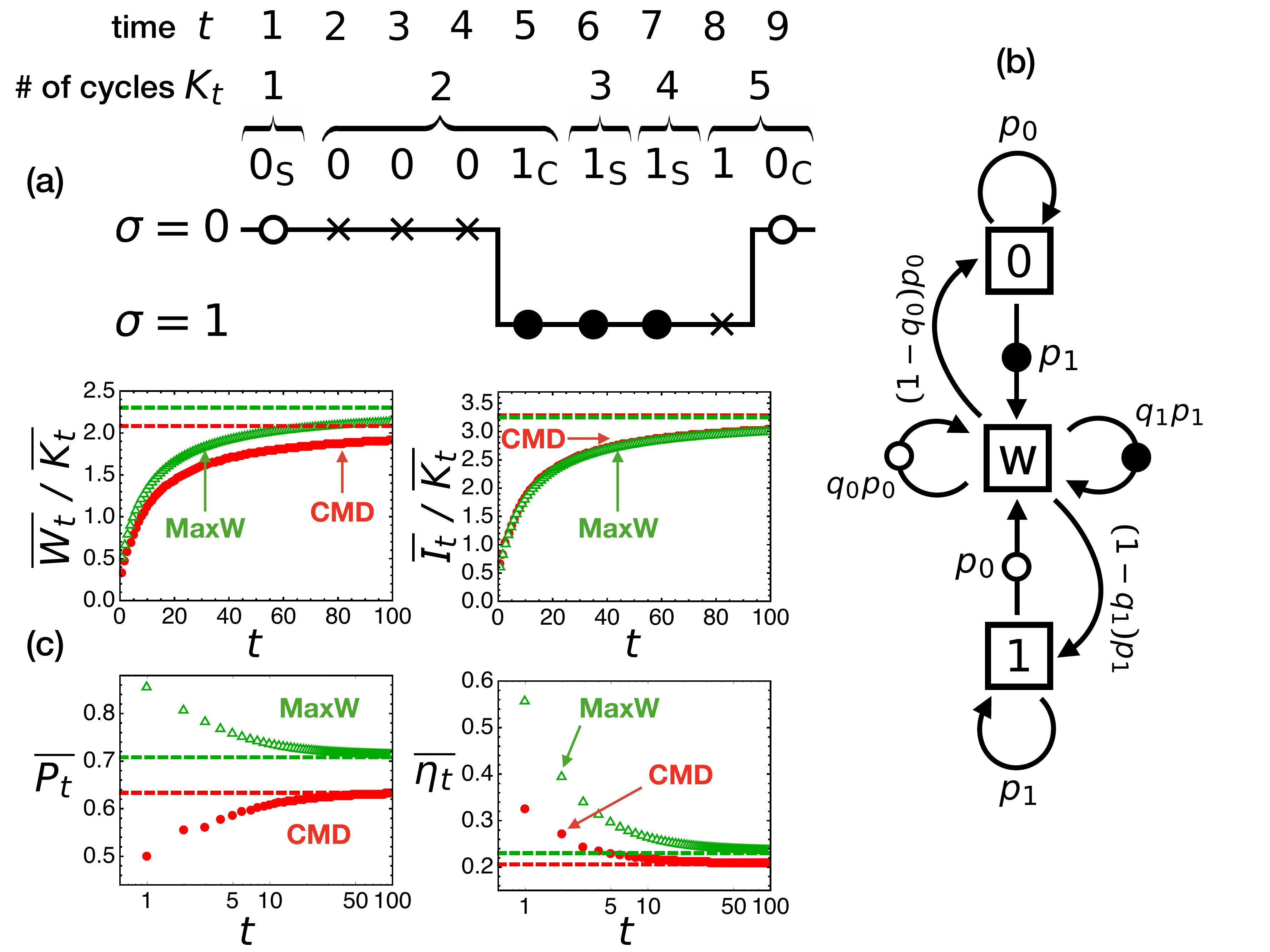} 
   \caption{{\bf  Power and efficiency at finite times: the 3-GCMD model.} (a) A series of repeated measurements terminating at a given finite time $t$ ($t=9$ in the illustration). Work is extracted at specific times (empty and filled circles) depending on the chosen protocol (SZ or CMD) and the measurement outcome ($\sigma=0,1$). Work is always extracted at the last measurement time $t$ even if a CMD-type cycle has not terminated. (b) Three-state (3-GCMD) model ($\sigma=$0,1,W) to calculate the work and information-content up to $t$. W denotes the third state corresponding to work extraction. (c) Average cycle-power and cycle-information (top) and average power and efficiency (bottom) versus $t$. Dashed lines are the $t\to\infty$ thermodynamic limit.}
   \label{fig:FIG4}
\end{figure}

\section{Enhanced power and efficiency by temporal correlations}
\label{sec:finitetime}
We can further assess the fluctuating nature of cycle power and efficiency by calculating the average power and efficiency measured over a finite time $t$ encompassing multiple work-extracting cycles. This defines a work-extracting engine that operates along consecutive cycles until time $t$, where the engine stops and work is extracted for the last time. To address this type of multi-cycle engine, we have extended the GCMD model of Fig.~\ref{fig:FIG1} to sequences of cycles, each cycle defined as before, i.e. a series of measurements terminating in a work extraction process, see Fig.~\ref{fig:FIG4}a. We calculate the average work, information, power, efficiency, and number of cycles over a finite time $t$ ($\overline{W}_t,\overline{I}_t,\overline{P}_t,\overline{\eta}_t,\overline{K}_t$). A sequence of measurements of time duration $t$ consists of a series of SZ-cycles and CMD-cycles that are selected with probabilities $q_\sigma$ and $1-q_\sigma$, respectively, depending on the measurement outcome $\sigma$ at the beginning of each cycle. The full sequence of measurements terminates at state $\sigma_t$ from which the last work $- \ln p_{\sigma_t}$ is extracted. The end state $\sigma_t$ determines three possible situations: it is the end of a CMD cycle, with $\sigma_t = 1-\sigma_{t-1}$; it is an SZ cycle; or it is the end time reached before a cycle of the CMD-type has completed, in which case $\sigma_t = \sigma_{t-1}$. 

Dynamics can be represented by three states, $\sigma=0,1$ (corresponding to mid-CMD cycle states), and an extra state $\sigma\equiv W$ for the work extraction steps. We denote this as the 3-GCMD model. The dynamics between the states are encoded in the three-state (discrete time) Markov network of Fig.~\ref{fig:FIG4}b. 
For instance, for t=9 a possible sequence in the 3-GCMD model could be (Fig.~\ref{fig:FIG4}b): $\lbrace \overbrace{0}_{\text{S}},\overbrace{0,0,0,1}_{\text{C}},\overbrace{1}_{\text{S}},\overbrace{1}_{\text{S}},\overbrace{1,0}_{\text{C}}\rbrace$. This sequence has $K_t=5$ cycles (three SZ -S- and two CMD -C- ones) and the work extracted equals $W_t=-2 \ln{p_0}-3 \ln{p_1}$. In the 3-GCMD model, the sequence reads $\lbrace W,0,0,0,W,W,W,1,W \rbrace$, and for calculating the average quantities ($\overline{W}_t,\overline{I}_t,\overline{P}_t,\overline{\eta}_t,\overline{K}_t$) it is not necessary to distinguish between the $W$ states (all that matters are transitions ending in $W$, Fig.~\ref{fig:FIG4}b). 

For $t\to\infty$ it is easy to prove that $\overline{P}_t\to P_{\rm th}$ and  $\overline{\eta}_t\to \eta_{\rm th}$, while $\overline{W}_t / \overline{K}_t 
\to W_{\rm th}$ and $\overline{I}_t / \overline{K}_t 
\to I_{\rm th}$, see Fig.~\ref{fig:FIG4}c. Significantly, in some cases, the 3-GCMD performs better at finite times than it does at long times: in the bottom panel Fig.~\ref{fig:FIG4}c we show that in the MaxW model $\overline{P}_t \geq P_{\rm th}$ for all times, while both in the CMD and MaxW the finite-time efficiency $\overline{\eta}_t \geq \eta_{\rm th}$ for all times. 

Throughout the paper, we neglected temporal correlations in the Demon state (i.e., $R\tau\gg 1$), making the statistics of all relevant quantities solvable for finite times in the 3-GCMD model. Interestingly, the intrinsic correlations of the 3-GCMD model (Fig.~\ref{fig:FIG4}b) make $\overline{I}_t / \overline{K}_t$ decrease faster than $\overline{W}_t / \overline{K}_t$ for lower $t$ (Fig.~\ref{fig:FIG4}c,top) enhancing efficiency (Figure \ref{fig:FIG4}c,bottom). Such correlations are absent in the Demon state for which we took $T_{\sigma'\sigma}=p_{\sigma'}$ ($T_{\sigma'\sigma}$ being the probability of measuring $\sigma'$ conditioned to measuring $\sigma$ at the previous time $\tau$). In general, $T_{\sigma'\sigma}$ also depends on $\sigma$, and such extra correlations should further increase efficiency. The role of temporal correlations in maximizing power and information-to-work efficiency has been considered in Ref.\cite{Admon2018} where a wall is repeatedly moved to rectify the motion of a diffusive colloidal particle against a flow. It would be interesting to extend the present analysis to finite $\tau$, and different stopping conditions \cite{neri2017,neri2019}, finding optimal regimes that maximize efficiency. Combining SZ and CMD-type protocols appears to be a promising route to develop improved protocols for information-to-energy conversion. These can be readily implemented in currently available experimental setups.

\subsection{Effect of finite-time work extraction and demon resetting}
\label{subsec:demonreset}
\begin{figure}[t!] 
   \centering
   \includegraphics[width=\columnwidth]{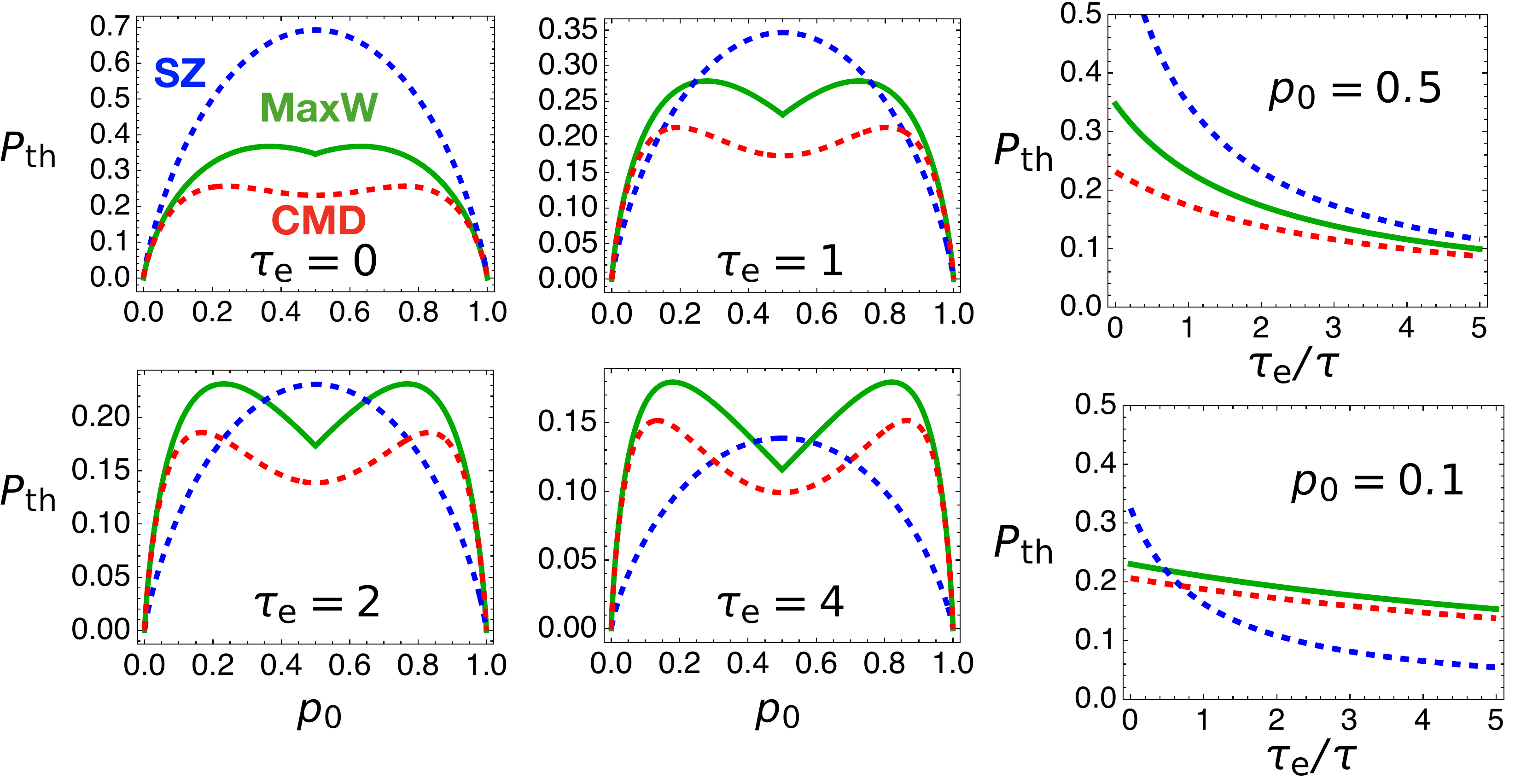} 
   \caption{{\bf  Effect of the finite-time $\tau_e$ for work extraction and demon resetting.} (a) Thermodynamic power $P_{\rm th}$ versus $p_0$ for the three models (SZ,CMD, MaxW) for different values of $\tau_e$ with $\tau=1$. (b) $P_{\rm th}$ versus $\tau_e/\tau$ for $p_0=0.5$ (top) and $p_0=0.1$ (bottom). GCMD models become more efficient than SZ for $\tau_e/\tau>1$.}
   \label{fig:FIG5}
\end{figure}

Throughout the paper, we assumed that the time of the work extraction and the resetting processes of the demon (erasure of the stored sequences) is negligible. This is a justified approximation in the CMD in the limit $R\tau\gg 1$ of largely-separated and uncorrelated measurements. Also, in this limit, redundancy of the stored sequences' information content is minimized, while the average extracted work is $\tau$-independent, making the information-to-work efficiency maximum for all GMCD cases. However, a finite-time duration $\tau_e$ for the combination of the work extraction and reset-erasure processes of the demon might change the thermodynamic power and information-to-work efficiency of generic GCMD models relative to SZ. Indeed, a finite time $\tau_e$ would be detrimental to the performance of the SZ where the total experimental time $t$ is increased by $K_t\tau_e$, where $K_t$ is the number of measurement cycles. In contrast, for the same $t$, the number of cycles of the CMD-type $K_t$ would be lower, and the average extracted work per cycle larger, increasing the thermodynamic power of the CMD relative to SZ. Therefore, for $\tau_e/\tau>1$, the thermodynamic power of SZ will decrease relative to the other models (CMD,MaxW,ConstW,TS). In figure \ref{fig:FIG5} we plot $P_{\rm th}$ as a function of $p_0$ (panel a) and $\tau_e/\tau$ (panel b) for the CMD (red), MaxW (green) and SZ (blue) models. GCMD models become thermodynamically more efficient than SZ for $\tau_e/\tau>1$. 

\section{Discussion}
\label{sec:discussion}
Is there any biological relevance to our model? There have been several applications of Maxwell demons to biological systems, particularly in the problem of signal transduction in  regulatory pathways with feedback loops. However, there is no work extraction in such cases but rather what has been termed transfer entropy and a second law relating the transfer entropy to the system's entropy production. It is no surprise there are no direct applications of Maxwell demons to work extraction processes in biological systems but rather entropy transfer calculations in regulatory pathways.

The GMCD model is a two-protocol information-to-energy conversion device.  Such a double route is akin to molecular folding models with two productive folding pathways: one pathway ends in a low-free energy molecular state while the other ends in a higher-free energy molecular state. Similarly, one can envision a regulatory process with two distinct feedback pathways with different transfer entropies per pathway. The applicability of the Maxwell demon to biology is still poorly understood. Whether Maxwell's demons are a direct product of biological evolution and life remains a long-debated and open question.

The CMD-type and SZ-type cycles might be put in analogy with the anabolic pathways of cell metabolism. In anabolic or biosynthetic pathways, high-energy complex molecules are built in a sequence of steps. Conversely, high-energy molecules are broken into smaller components in catabolic or degradation pathways. In this analogy, the high information-content sequences (multiple-bit sequences) stored in the CMD might be seen as the equivalent of the high-energy molecules produced in the anabolic pathway. Instead, the low information-content sequences (single-bit sequences) in the SZ, are analogous to the much lower free energy molecules that can be produced in biosynthetic reactions with a fixed number of steps. A typical example of an anabolic reaction of the SZ-type is glucose production from carbon dioxide in photosynthetic cells. Here the SZ-type cycle is envisioned as a well-defined reaction that produces a molecule (glucose) with a well-defined amount of energy.
In contrast, an anabolic reaction of the CMD-type might be realized in the biosynthesis of a polypeptide chain. In protein synthesis, the amino acids are sequentially linked in a three-step process (initiation, elongation, and termination) where termination occurs when a specific condition (stop codon) is met. The analogy between the CMD and protein synthesis lies in the proteins of variable length and free energy that can be assembled during ribosomal translation.  This is analogous to the varying information-content of the (multiple bits) sequences in the CMD. The interest of the GCMD model is that it allows for two information-content pathways, reminiscent of the different biosynthetic pathways of anabolism. An efficient metabolism requires synthesizing many molecules of variable length and free energy, each specific molecular type for a given metabolic step. Analogously, the GMCD model considers two information-to-energy conversion pathways (SZ and CMD) that generate bit sequences of variable information-content (single-bit sequences for SZ versus multiple-bit sequences for CMD).

\section{Conclusions}
\label{sec:conclusions}
An important quest in the field of thermodynamics of information is to search for protocols that maximize the information-to-energy conversion efficiency, a quantity that has been hypothesized to be optimized in small biological systems. Indeed, the large efficiency of many molecular motors and enzymes is remarkable. Light harvesting is an example of how the energy of a single light photon absorbed by chlorophyll powers single electron transfer reactions with almost 100$\%$ quantum efficiency. In our paper, we digress about neuronal transmission as an example of a work extraction cycle of three phases under information feedback: information storage (cell depolarization), state check (potential threshold), and energy release (current spike). It is argued that these activity cycles resemble more the continuous version of the MD (CMD) than the Szilard protocol (SZ). Indeed, no state-check condition is present in SZ, which is equivalent to a hypothetical neuronal transmission activity without threshold-driven spiking. There is no irrefutable evidence that efficiency is an  optimized quantity in small biological systems, although it is clear the high efficiency achieved by biological processes when compared to equivalent man-made machines. 

A general result of our study is the larger thermodynamic efficiency of the SZ relative to the different variants of the GMCD with a CMD component ($q_\sigma<1 0$). However, this holds if we assume a negligible time for the work extraction and erasure process (Sec.\ref{subsec:demonreset}). Moreover, the high thermodynamic efficiency of SZ comes at the price of a bounded extracted work per cycle ($<k_BT\log 2\sim 0.69k_BT$, Figure \ref{fig:FIG2}a). In contrast, the average work per cycle is much larger in the CMD, being unbounded in the limit $P_0\to 0$ and $P_0\to 1$.  The lower thermodynamic efficiency of the CMD stems from the redundancy of the information-content encoded in the multiple-bits stored sequences. Such a redundancy led us in Sec.\ref{subsec:physical} to classify SZ as a low-risk strategy (low both average work and information-content per cycle) compared to CMD (both large average work and information-content per cycle). The large thermodynamic power for SZ is concurrent with the low work per cycle ($<k_BT\log 2\sim 0.69k_BT$), which does not even reach $1k_B T$, poses serious limitations to SZ to perform arbitrary information-to-energy conversion tasks. From the standpoint of biological systems, it suggests that information-to-work efficiency is more relevant than thermodynamic power. The limited amount of extracted work per cycle does not cloud the CMD, a better model for biological processes in an information-redundant world.

Most results presented in this paper are purely mathematical. The message we tried to convey is that CMD realizations store words of multiple-bit sequences that can be converted into large amounts of work. The Szilard engine is the simplest single-bit information-to-energy conversion machine with limited average work and information-content per cycle, yet there is room for designing smart information-to-work protocols in systems with a finite number of states $\Omega$ that can extract more work than just $k_BT\log \Omega$. This might be achieved with machines that can store multiple-bit sequences of large information-content. Plausibly, information-content is also the main trait of the living entities generated during eons of natural evolution in an information-redundant world.

\acknowledgements JPG acknowledges
financial support from EPSRC Grant no.\ EP/R04421X/1. JPG is grateful to All Souls College, Oxford, for support through a Visiting Fellowship during the latter stages of this work. FR acknowledges
support from  ICREA Academia 2018, Spanish Research Council Grant FIS2016-80458-P, and PID2019-111148GB-I00.

\bibliographystyle{apsrev4-1}
\bibliography{GCMD}

\end{document}